# Influence of dissociative recombination on the LTE of argon high-frequency plasmas at atmospheric pressure

A. Sáinz[1], J. Margot[2], M. C. García[1], M. D. Calzada[1]

[1] *Grupo de Espectroscopía de Plasmas, Universidad de Córdoba. Edificio C-2. Campus Rabanales. 14071 Córdoba (Spain)*
[2] *Groupe de Physique des Plasmas, Université de Montréal, C. P. 6128, Succ. Centre-ville, Montréal, Quebec, H3C3J7, (Canada).*

## 1. Introduction

Current literature shows that surface-wave sustained discharges (SWDs) can advantageously be used as sources of excitation in mass spectrometry and atomic emission spectrometry to determine the composition of analytical samples [1-2]. In particular, SWDs are most advantageous than Inductively Coupled Plasmas (ICPs) for the excitation of halogens and other nonmetals that are not readily accessible to ICP detection with the adequate sensitivity.

In order to improve and optimize the characteristics of SWDs for applications, it is necessary to achieve a good understanding of the discharge as a function of the experimental conditions. The crucial parameters characterizing the discharge and therefore influencing its analytical performances are the electron density ($n_e$) and temperature ($T_e$), as well as the gas temperature ($T_g$). Because of their relatively high density, other species such as atoms ($n(p)$) and ions ($n^+$) are also expected to play an important role in the discharge kinetics.

At atmospheric pressure, the electron temperature can often be difficult to determine experimentally and, for this reason, its value is frequently speculated. Assuming the plasma to be in Local Thermodynamical Equilibrium, the distribution of the population of the excited levels obeys a Saha-Boltzmann distribution with an excitation temperature identical to the electron temperature. In this equilibrium situation, the production and loss of the electrons at a microscopic level is controlled by the Saha balance and the population distribution of the excited levels obeys the Saha-Boltzmann distribution. However, at atmospheric pressure, molecular recombination channels may also play an important role in the plasma kinetics. In such a case, equality between excitation and electron temperature is questionable.

Molecular ions principally form via atom-assisted association:

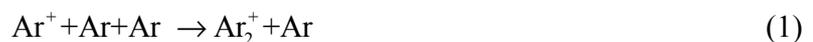
$$Ar^+ + Ar + Ar \rightarrow Ar_2^+ + Ar \qquad (1)$$

and are lost through the dissociative recombination process:

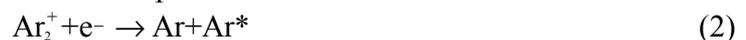
$$Ar_2^+ + e^- \rightarrow Ar + Ar^* \qquad (2)$$

where the asterisk denotes a possible excitation state of one of the resulting atoms.

The present knowledge on the distribution of dissociative products and its influence over the plasma excitation kinetics is quite incomplete. In the case of argon, it is usually assumed that the most important dissociative recombination channel of dimer ions yields one argon atom in the 4p state and the other one in the ground state [3]. However, it has been recently found that dissociative recombination results in a significant number of excited atoms in the 4s state and a non negligible fraction of atoms in the ground state [4-5].

In the present work we present a few preliminary results from a collisional-radiative (CR) model that we developed for an argon microwave (2.45 GHz) plasma at atmospheric pressure. This model aims to investigate the influence of dissociative recombination products on the Saha-Boltzmann plasma equilibrium. The model is tested through comparison with experimental results obtained in an argon plasma column generated by a traveling electromagnetic surface-wave in a cylindrical tube open to the air [7-9]. In the traveling wave regime, such plasmas are characterized by a plasma density and electron temperature that decrease as one moves from the wave launcher region (where power flux is maximum) to the end of plasma column (where power flux comes to zero). Therefore, with only one plasma column, it is possible to perform a parametric investigation of the plasma, which is quite suitable for model testing.

## 2. Collisional model
### 2.1 Kinetics of the model

The scheme of argon energy levels is based on the model of Vlček [6] considering the atomic levels up to 7p'. The main processes included in the model are:
(i)   Stepwise excitation and de-excitation by electron impact,
(ii)  Stepwise ionization,
(iii) Three-body recombination,
(iv)  Radiation including trapping,
(v)   Excitation and de-excitation by atom impact,
(vi)  Diffusion of the charged particles (ambipolar),
(vii) Quenching and diffusion of the metastable states,
(viii) Dissociative recombination,
(ix)  Three-body interaction.

### 2.2 Conditions for the simulations
All the simulations were performed using with the following conditions: $T_g$ = 1500 K, P = 760 Torr, a = 0.5 mm (radius of the tube), which correspond to the experimental observations reported by Calzada et al. [7], Garcia et al. [8] and Santiago [9]. The electron density values along the plasma column were taken from the work of Calzada et al [7].

## 3. Results and discussion
### 3.1 Influence of the dissociative recombination products on the excited levels populations
As mentioned earlier, some authors [3, 10-11] consider that the most important products from dissociative recombination yield one argon atom in the ground state (p = 1) and another one in the 4p excited level. Other researchers reported that the process should rather yield atoms in the 4s state and in the ground state [4-5]. These cases correspond to the following reactions

$$Ar_2^+ + e^- \rightarrow Ar(1) + Ar(4p) \qquad (3)$$
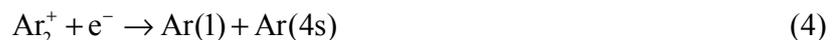
$$Ar_2^+ + e^- \rightarrow Ar(1) + Ar(4s) \qquad (4)$$

Using our collisional-radiative model, we calculated the excited level populations considering the DR products given either by Eq.(3) or by (4). The results were compared to the populations experimentally determined. Figure 1 shows that the reaction products are crucial for the populations of excited levels. Actually, a better agreement between theory and experiment is achieved when considering Eq. (3) rather and Eq. (4). This indicates that DR produces a large number of argon atoms in 4s state.

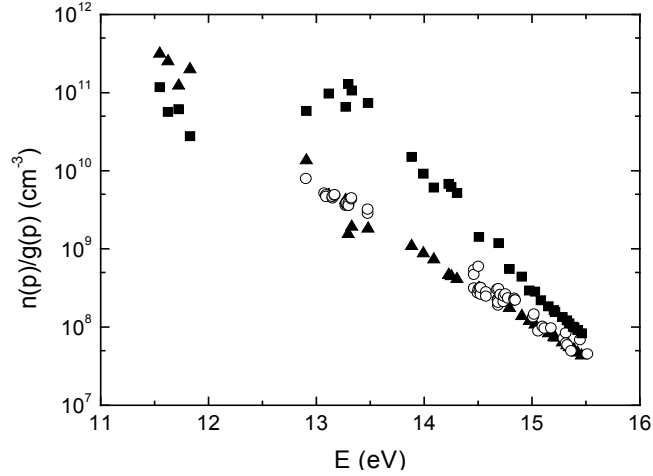

**Figure 1:** Population density of excited atoms calculated assuming that one of the dissociative recombination products is the 4s state (triangles) and the 4p state (squares). The calculations were performed for $T_g$=1500 K, $T_e$=5600 K and $n_e$=4 × $10^{14}$ cm$^{-3}$. The results are compared with available experimental data.

In order to estimate the weight of each possible product resulting from DR, we have performed calculations by considering that the reaction could produce argon atoms in ground state as well as in the 4s, 3d, 4p and 5p states with fractions $\chi(1)$, $\chi(4s)$, $\chi(3d)$, $\chi(4p)$ and $\chi(5p)$, respectively. The equation system was closed by specifying two conditions: 1) the sum of the $\chi$ fractions is equal to unity, and 2) $\chi(3d)/\chi(4p) \approx 0.7$ as stated in reference [12]. It was thus found that the best fit between model and experiments occurs for $\chi(1) = 0.25$, $\chi(4s) = 0.70$, $\chi(4p) = 0.02$, $\chi(3d) = 0.02$, while $\chi(5p)$ is negligible. Thus, it appears that dissociative recombination predominantly populates the 4s levels and the ground state, in agreement with the experimental results obtained by Ramos et al. [4-5].

### 3.2. Influence of $Ar_2^+$ processes in the plasma thermodynamic equilibrium degree

In order to quantitatively determine the departure from local Saha Equilibrium, we calculated the parameter $b(p) = n(p)/n^s(p)$, where $n(p)$ and $n^s(p)$ denote the population density of atoms in the level p and the corresponding Saha population and p is the effective quantum number ($p = \zeta(E_H/E_{ip})^{1/2}$, where $\zeta$ is the core charge of the atom, $E_H$ the hydrogen ionization potential, and $E_{ip}$ the ionization potential of the atom in the level p). Those levels for which b = 1 are in Local Saha Equilibrium.

Figures 2a and 2b compare the calculated b(p) values to those measured experimentally for different z positions along the plasma column, i.e. various ($n_e$, $T_g$) sets. Calculations were carried out for two situations: one considering dissociative recombination and another one where it is neglected. It is observed from Fig. 2a and 2b that DR strongly influences the population of the levels, specially those of lower energy (p < 3). Actually, considering DR reduces departure from equilibrium. It can also be seen that the experimental data obtained by Calzada et al. [6] fall close to the theoretical results obtained with DR. Figures 3a and 3b further show that b(p) increases with p until it reaches unity. This occurs for p ~ 4 at low plasma density (z = 2 cm) and for p ~ 2 at higher density (z = 8 cm). In this later case, almost all levels are in equilibrium except the group of the 4s levels. Thus the plasma is closer to equilibrium as $n_e$ increases, which is no real surprise since LTE is more likely to occur in denser plasmas. In all cases, the higher levels (close to the ionization limit) appear to be in equilibrium. This allows assuming that the excitation temperature $T_{exc}$ determined from the upper levels in the atomic system in the Boltzmann-plot is equal to $T_e$.

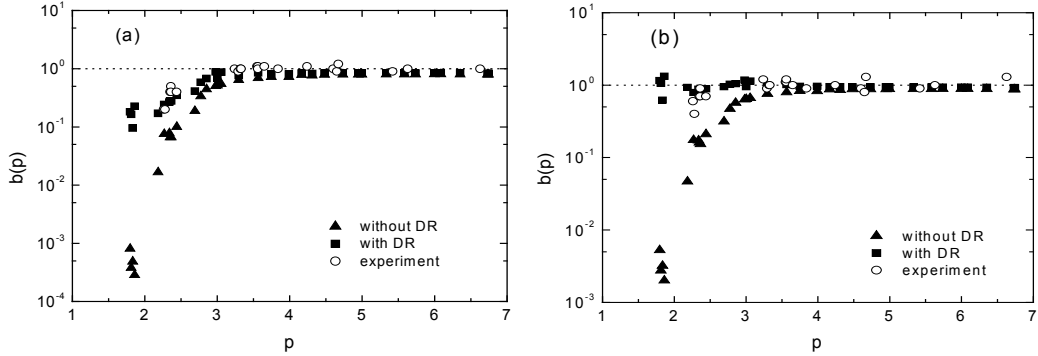

**Figure 2:** Departure from local Saha equilibrium for (a) $T_g$=1300 K, $T_e$=4500 K and $n_e$=3.2× $10^{14}$ cm$^{-3}$ and (b) $T_g$=1500 K, $T_e$=6000 K and $n_e$=4.5× $10^{14}$ cm$^{-3}$

From Figs. 2a and 2b, the plasma clearly presents a recombining behavior since b(p) is almost everywhere smaller than 1. As just mentioned the 4s levels are those that show the most significant departure from equilibrium. It is particularly striking for the $^3P_2$ metastable population as illustrated in Fig. 3. This figure shows the calculated values of b(p) for the $^3P_2$ level using as input parameters to the model those experimentally observed at various axial positions along a surface-wave plasma column. For comparison, the experimental data obtained by Santiago [12] using a self-absorption method are also shown.

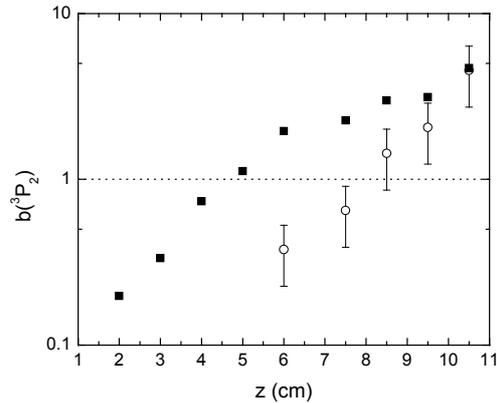

**Figure 3:** Departure from local Saha equilibrium for the $^3P_2$ metastable population. The calculations were performed for input parameters ($n_e$ and $T_e$) that correspond to various positions z along a surface-wave plasma column.

As can be seen, both model and experiment show that for this level, the discharge exhibits a recombining-to-ionizing transition as electron density increases, i.e, when moving from the end of plasma column (smaller z values) to larger z positions. Despite the small discrepancy between theory and experiment on the z position at which transition occurs, it is clear that that the discharge is characterized by two zones with different internal kinetics. The first one (higher z values) corresponds to higher electron density. Ions and electrons are continuously created, likely from the metastable levels, and creation is balanced by recombination that maintains the $n_e$ value constant under steady-state conditions. In the second zone (lower z values, i.e. lower electron density values), the 4s and 4p levels are underpopulated with respect to the corresponding Saha-Boltzmann populations because of the gradual loss of energy by the surface wave that favors recombination.

## 4. Acknowledgments

This work was supported by the Spanish Ministry of Science and Technology under grant no. FTN2002-02595 and the European Economic Community in the frame-work of the Feder Funds.